\documentstyle[epsfig,aps]{revtex}
\draft	

\begin{document}
\title{Dynamics of Ordering of Isotropic Magnets}
\author{Jayajit Das\cite{JAY}} 

\address{Institute of Mathematical Sciences, CIT Campus, Taramani, Madras 
600113, India}

\author{and}

\author{Madan Rao\cite{MAD}}

\address{Raman Research Institute, C.V. Raman Avenue,
Sadashivanagar, Bangalore 560080, 
India}

\date{\today}

\maketitle

\begin{abstract}

We study the dynamics of ordering of the nonconserved and conserved
Heisenberg magnet. The dynamics consists of two parts $-$ an irreversible
dissipation into a heat bath and a reversible precession induced by a
torque due to the local molecular field. For quenches both to $T=0$ and
$T=T_c$, we show that the torque is irrelevant when the dynamics is
nonconserved but relevant when the dynamics is conserved and is governed
by a new nontrivial fixed point.

\end{abstract}

\hspace{1.5 cm} Keywords : Heisenberg
magnet, Langevin equation, Dynamical scaling, Multiscaling.
\vskip 0.1 in 

\section{Introduction}

Most interacting systems like magnets and binary fluids when quenched from
the disordered phase to zero temperature, evolve extremely slowly to
establish order, primarily because of the slow annealing of the interfaces
(defects) separating the competing domains. It is seen that at late times,
the system organizes itself into a self similar spatial distribution of
domains characterised by a single diverging length scale which typically
grows algebraically in time $L(t) \sim t^{1/z}$. The equal-time order
parameter correlation function $C(r, t) \equiv \langle {\vec \phi}(r, t)
\cdot {\vec \phi}(0,t) \rangle$ is a measure of the spatial distribution
of the domains, and at late times is found to behave as $f(r/L(t))$, where
$L(t)$ is the distance between defects.  The autocorrelation function,
$C(0, t_{1}=0, t_{2}) \equiv \langle \,{\vec \phi}(0, 0) \cdot {\vec
\phi}(0, t_2) \rangle$, is a measure of the memory of the initial
configurations, and decays at late times as $L(t_{2})^{-\lambda}$. The
independent scaling exponents $z$ and $\lambda$ and the scaling function
$f(x)$ characterise the dynamical universality classes at the zero
temperature fixed point \cite{BRAY}.

There has been a trend in recent years to compare the theories of phase
ordering dynamics with numerical simulations of Langevin equations.
Comparison with experimental systems, such as magnets, binary fluids or
binary alloys have to take into account the various `real' features that
might be relevant to its late time dynamics. For instance, theories of
binary fluids have to include effects of hydrodynamics, while those of
binary alloys have to incorporate elastic and hydrodynamic effects. In the
same vein, any comparison with the dynamics in real magnets has to include
the effects of the torque induced by the local molecular field
\cite{DAS1,DAS2}.

\section{Phase Ordering Dynamics\,: Quenches to $T = 0$}

The spins $\phi_{\alpha}$ ($\alpha = 1,2,3$) in a Heisenberg
ferromagnet in three dimensions experience a torque from the joint
action of the external field (if present) and the local molecular
field. In response the spins precess with a Larmour frequency
$\Omega_L$ about the total magnetic field. Coupling to various faster
degrees of freedom like phonons, electrons and magnons causes
dissipation and an eventual relaxation towards
equilibrium. 

The equations of motion in dimensionless variables is given by
\cite{DAS1},
\begin{equation}
\frac{\partial \vec{\phi}}{\partial t} =
-(-i\nabla)^{\mu}\left(\nabla^2 \vec{\phi} \, + \,  \vec{\phi} - \,
\left(\vec{\phi} \cdot \vec{\phi}\right)\, \vec{\phi}\right) \, + \,g
\, \left(\vec{\phi} \times
\nabla^2\vec{\phi}\right)\, .
\label{eq:dynamic}
\end{equation}
The exponent $\mu$ in the above equation takes the value $0$ when the
magnetisation is not conserved (NCOP) and $2$ when it is
conserved (COP).
The dimensionless parameter $g \sim \Omega_{L}/\Gamma$ is
the ratio of the precession frequency to the relaxation rate, which
is in the range $g \sim 10^{-3} - 10$ for typical ferromagnets.

\subsection{Langevin Simulation} 

We prepare the system in the paramagnetic phase and quench to zero
temperature. We study the time evolution of the spin configurations as
they evolve according to Eq.\ (\ref{eq:dynamic}). We calculate the
spatially averaged equal time correlator $C({\bf r} ,t )$ and the
autocorrelator $C({\bf 0}, t_0=0, t)$, both averaged over the random
initial conditions. We compute the scaling exponents $z$ and $\lambda$ and
the scaling function $f(r/L(t))$ as defined above.  $L(t)$, a measure of
the distance between defects, is extracted both from $C({\bf r},t)$ and
from the scaling form of the energy density, $ \varepsilon = \frac{1}{V}\,
\int d {\bf r} \,\langle \,(\,\nabla \phi({\bf r}, t)\,)^2 \,\rangle$.

We discretize Eq.\ (\ref{eq:dynamic}) on a simple cubic lattice of size
$N$ ranging from $40^3$ to $60^3$, adopting an Euler scheme for the
derivatives \cite{SR} with periodic boundary conditions. The space and
time intervals have been chosen to be $\triangle x = 3$, $\triangle t =
0.01$ (NCOP) and $\triangle x = 2.5$, $\triangle t = 0.20 $
(COP).  Further details of the simulation may be found in
\cite{DAS1,DAS2}.

\subsubsection{Non-Conserved Case}

We will show that the torque is irrelevant in $d=3$, which means that the
asymptotic values of $z$, $\lambda$ and the scaling form $f(x)$ remain
unchanged and independent of $g$. The correlation functions have been
computed for $g$ ranging from $0$ to $5$. The dynamical exponent is found
to be unchanged from $z=2$ (within statistical errors) and independent of
$g$ \cite{DAS1}. Likewise the autocorrelation exponent is
unchanged from its $g=0$ value and equal to $\lambda \approx
1.52$ independent of $g$. Note that the
numerical determination of $\lambda$ is subject to large
errors\cite{DAS1,YRD}, and so we have to go to very late times and hence
large system sizes.

\begin{figure}
\centerline{\epsfig{figure=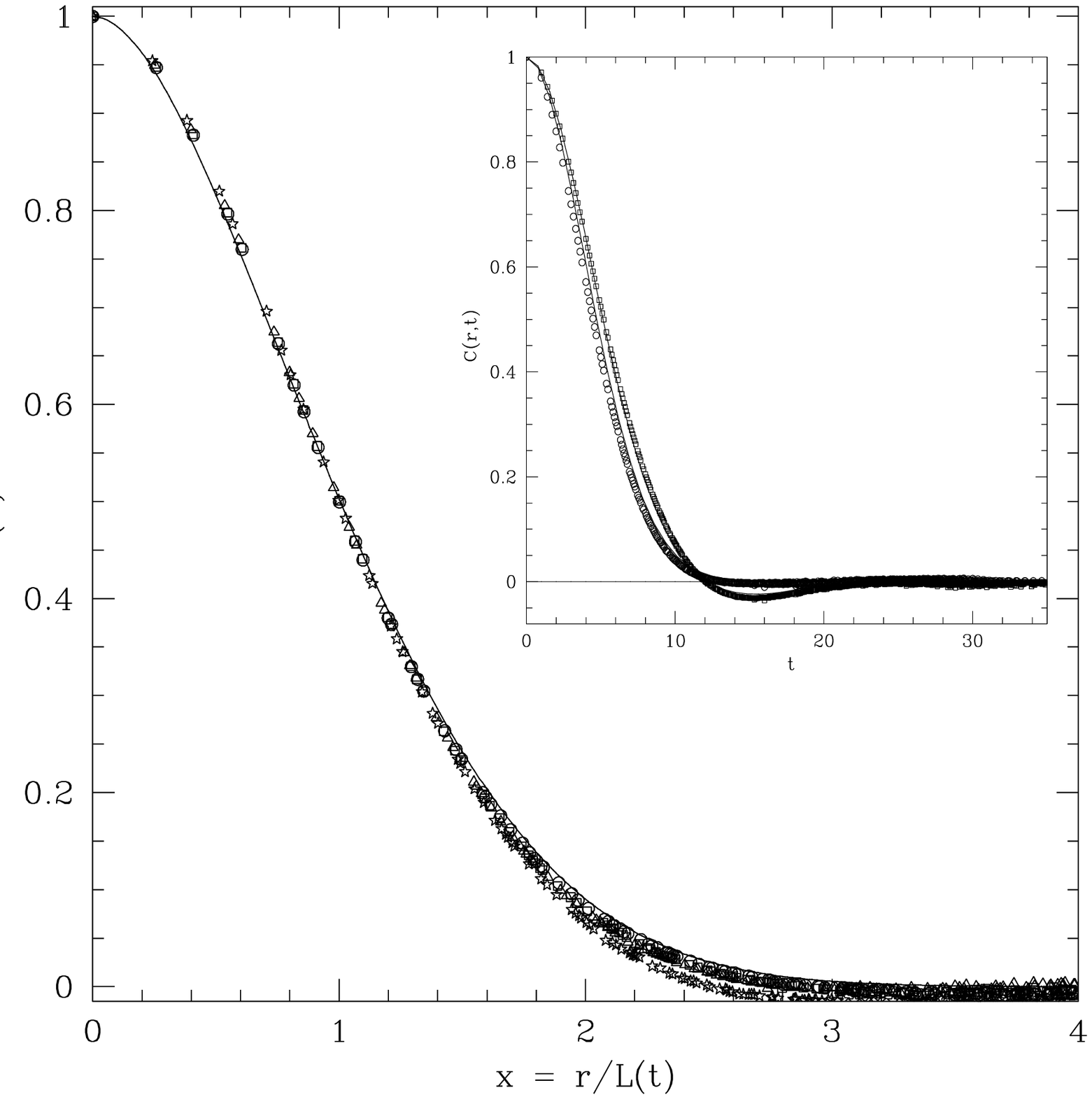,width=7.0cm,height=7.0cm}}
\end{figure} 
\noindent
Fig. 1. Scaling plot of $C(r,t)\equiv f(x)$ versus $x\equiv r/L(t)$ for $g =
0(\circ),\, 0.5(\triangle),\,2(\ast)$. The solid line through the data
points is a fit to the BPT function \cite{BPT}. Inset compares $C({\bf
r},t)$ for $g=1(\circ)$ with $g=5(\Box)$ at late times. The solid line
through these data points comes from a preasymptotic analysis (see below)
of $C(r,t)$ \cite{DAS1}.  
\vspace{1 cm}

The scaling function $f(x)$ is also unchanged and independent of $g$. This
is clearly apparent for $g = 0,\,0.5,\,1,\,2$ (Fig.\ 1). However at larger
values of $g$, the correlation function crosses zero at large $r$, dips
through a minimum, and then asymptotically goes to zero (of course $\int
C(r,t) > 0$). It would appear that $C(r/L(t))$ for $g=5\,$ is qualitatively
different from the scaling function for $g=0$. However we note that the
dip decreases very slowly with increasing time suggesting it might
disappear at late times. Patience confirms this for intermediate values of
$g$ (between $2$ and $5$). When $g=5$, finite size effects prevent the
system from exploring its true asymptotic regime, when the order parameter
field has totally relaxed with respect to defect cores. Pre-asymptotic
configurations typically consist of spin wave excitations interspersed
between slowly moving defects separated by a distance $L(t) \gg \xi$, the
size of the defect core. Decomposing $\vec \phi$ into a singular (defect)
part $\vec \phi_{sing}$ and a smooth (spin wave) part $\vec \phi_{sm}$
gives the preasymptotic correlation function three contributions $-$
$C_{sing}\equiv \langle \vec\phi_{sing}(0,t)\cdot \vec\phi_{sing}({\bf
r},t)\rangle$ (defect contributions), $C_{sm} \equiv\langle
\vec\phi_{sm}(0,t)\cdot \vec\phi_{sm}({\bf r},t)\rangle$ (spin wave
contributions) and $C_{scat}\equiv \langle \vec\phi_{sm}(0,t)\cdot
\vec\phi_{sing}({\bf r},t)\rangle$ (scattering of spin waves off slowly
moving defects). The defect contribution $C_{sing}$ takes the form given
in \cite{BPT}, while the spin wave contribution $C_{sm}$ can be evaluated
perturbatively \cite{DAS1}. Even without including the scattering between
spin waves and defects, we find that the total correlation function has a
dip when $g\neq0$ as shown by the solid line in the inset of Fig.\ 1. The
long lifetime of the dip arises because of the slow relaxation of the spin
waves as they scatter off the slow moving defects.

\subsubsection{Conserved Case}

When the dynamics is conserved, we show that the the torque
$g$ drives the system to a new fixed point characterised by a different
value of $z$, $\lambda$ and the scaling function $f(x)$. Moreover these
quantities are independent of the value of $g$ as long as it is $g>0$.
This crossover is described by a crossover exponent and a crossover
function.

Figure 2 shows that the scaling function for $g=0$ is very different from
those for $g>0$\, and the $g>0$ scaling functions do not depend on the
value of $g$.  The $z$ exponent crosses over from $z=4$ (its value at
$g=0$) when $t<t_c(g)$ to $z=2$ when $t>t_c(g)$ where $t_c(g)$ is the
crossover time which decreases with increasing $g$ (Fig. 2(inset)).
Likewise the autocorrelation exponent $\lambda$ crosses over from 
$\lambda\approx 2.2$ when $g=0$ to $\lambda \approx 5.15$ when 
$g>0$ \cite{DAS2}. 

\begin{figure}
\centerline{\epsfig{figure=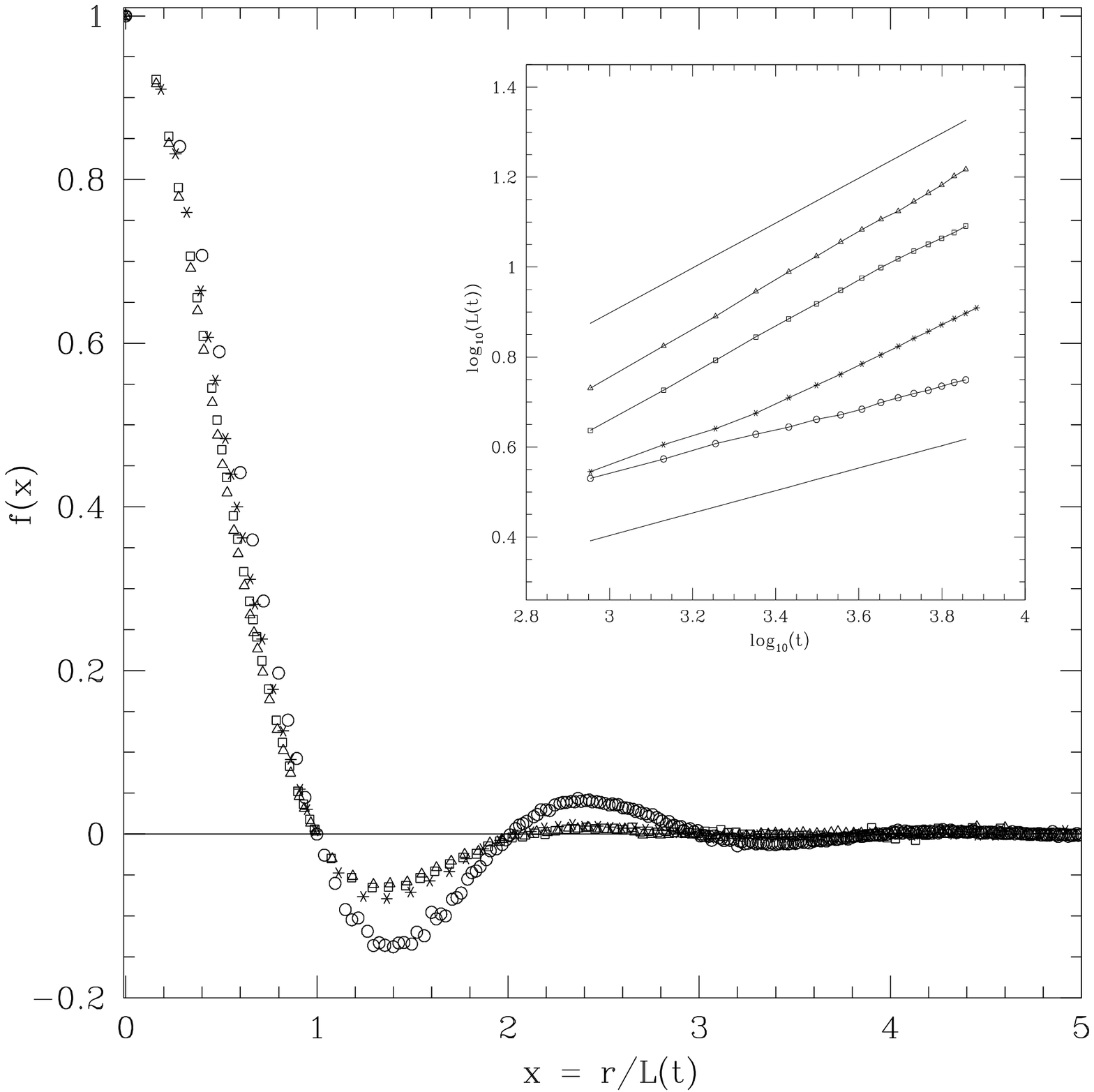,width=7.0cm,height=7.0cm}}
\end{figure}
\noindent
Fig.\ 2. Scaling plot of $C({\bf r},t)$ comparing $g=0(\circ)$ with $g>0$
($g=0.1(\ast),\,0.3(\triangle)\,0.5(\Box)$. Inset shows
log-log plot of $L(t)$ versus $t$.  At $g=0\,(\circ)$, $z=4$ (line of
slope $0.25$ drawn for comparison). At other values of $g\,
(g=0.1(\ast),\,0.3(\Box),\, 0.5(\triangle))$, $z$ crosses over from
$4$ to $2$ (line of slope $0.5$ drawn for comparison). 

\vspace{1 cm}

The crossover may be understood from a simple scaling argument
\cite{DAS2}. On restoring appropriate dimensions, the
dynamical equation Eq.(\ref{eq:dynamic}) can be rewritten as a
continuity equation, $\partial{\vec\phi({\bf r},t)}/{\partial t } = -
{\bf \nabla }\cdot{\vec j }$ where the ``spin current''

\begin{equation}
 \vec{j}_{\alpha} = -\Gamma\left({\bf \nabla }\frac{\delta F[\vec
\phi]}{\delta \phi_{\alpha}} +
\frac{\Omega}{\Gamma}\epsilon_{\alpha\beta\gamma}\phi_{\beta}\nabla\phi_{
\gamma}\right) \, . 
\label{eq:current} 
\end{equation} 
Using a dimensional analysis where we replace $j_{\alpha}$ by the
`velocity' $dL/dt$, we find

\begin{equation} 
\frac{dL}{dt} = \Gamma\frac{\sigma}{L^3} + \Omega\frac{\sigma
M_{0}}{L}\, ,
\label{eq:dimension} 
\end{equation}
where the parameters $M_0$, $\sigma$ and $\Gamma^{-1}$ are the
equilibrium magnetisation, surface tension and spin mobility
respectively. Beyond a crossover time $t_c(g) \sim
(\Gamma/M_{0} \Omega)^2 \sim 1/g^2$, simple dimension counting shows
that the dynamics crosses over from $z = 4$ to $z = 2$ in conformity
with our numerical simulation. Our numerics supports a scaling form 
$L(t,g)=t^{1/4}s(tg^{\phi})$ with $\phi \approx 1.7$, valid for all
$g$ \cite{DAS2}. 

\subsection{Numerical Tests of the Mazenko Closure Scheme}

There is a class of approximate theories  based on the gaussian
closure approximation \cite{MAZENKO,BRAY} that is supposed to provide
fairly successful predictions for the forms of the correlation
functions. We will show that the gaussian closure approximation
works fairly well for the nonconserved Heisenberg model with $g \geq 0$
and leads to conclusions similar to the ones
described in the previous section. In contrast, the use of the
gaussian closure in conserved models leads to sharp inconsistencies.

The gaussian closure method consists of trading the order parameter
${\vec \phi} ({\bf r},t)$ which is singular at defect sites, for an
everywhere smooth field ${\vec m}({\bf r},t)$, defined by a nonlinear
transformation, $ {\vec \phi}({\bf r}, t) = {\vec \sigma}\left({\vec
m}({\bf r} ,t)\right)\, $. An appropriate choice for this nonlinear
function ${\vec  \sigma}$ is an equilibrium defect profile,
\begin{equation}
{\vec \sigma}\left({\vec m}({\bf r},t)\right) = \frac{{\vec m}({\bf
r}, t  )} {\vert{\vec m}({\bf r},t) \vert}\,g({\vert{\vec m}\vert})\,,
\label{eq:hedgehog}
\end{equation}
where $g(0)=0$ and $g(\infty)=1$. With this choice, $\vert \vec
m\vert$ has the interpretation of being the distance away from a
defect core.   Correlation functions are calculated making the single
assumption that each component of ${\vec m}({\bf r},t)$ is an
independent gaussian field with zero mean at all times
\cite{MAZENKO,BRAY}. 

To check whether this is a good approximation we compute $\vec{\phi}$
numerically and then determine ${\vec m}({\bf r},t)$ by inverting Eq.\
(\ref{eq:hedgehog}).  We then calculate both the single point
probability distribution $P({\vec m}({\bf r},t))$ and the joint
probability distribution in the scaling regime and compare
with the gaussian assumption of Mazenko. These computations require a
lot of averaging over initial configurations to obtain good
statistics. 

\subsubsection{Nonconserved Case}

Figure 3 shows the scaling plots of $P(m_1({\bf r},t))$ ($m_1$ is a
component of ${\vec m}$) at $g=0$ and $g=1$. Since $m$ has dimensions
of length, the appropriate scaling variable is $m_1/L(t)$
\cite{DAS1}. The scaled distribution $P(m_1)$ is seen to be the same
for $g=0$ and $g=1$, suggesting that it is unchanged on addition of
the torque. We also find that it is independent of $g$. 

The scaled distribution $P(m_1)$ closely resembles a gaussian at late
times (this is confirmed by a more detailed analysis \cite{DAS1}).
The calculated joint probability distribution also  agrees reasonably
well with the Mazenko assumption \cite{DAS1}.  These findings are
consistent with a similar analysis done on a scalar order parameter
\cite{CHUCK}. 

\begin{figure}
\centerline{\epsfig{figure=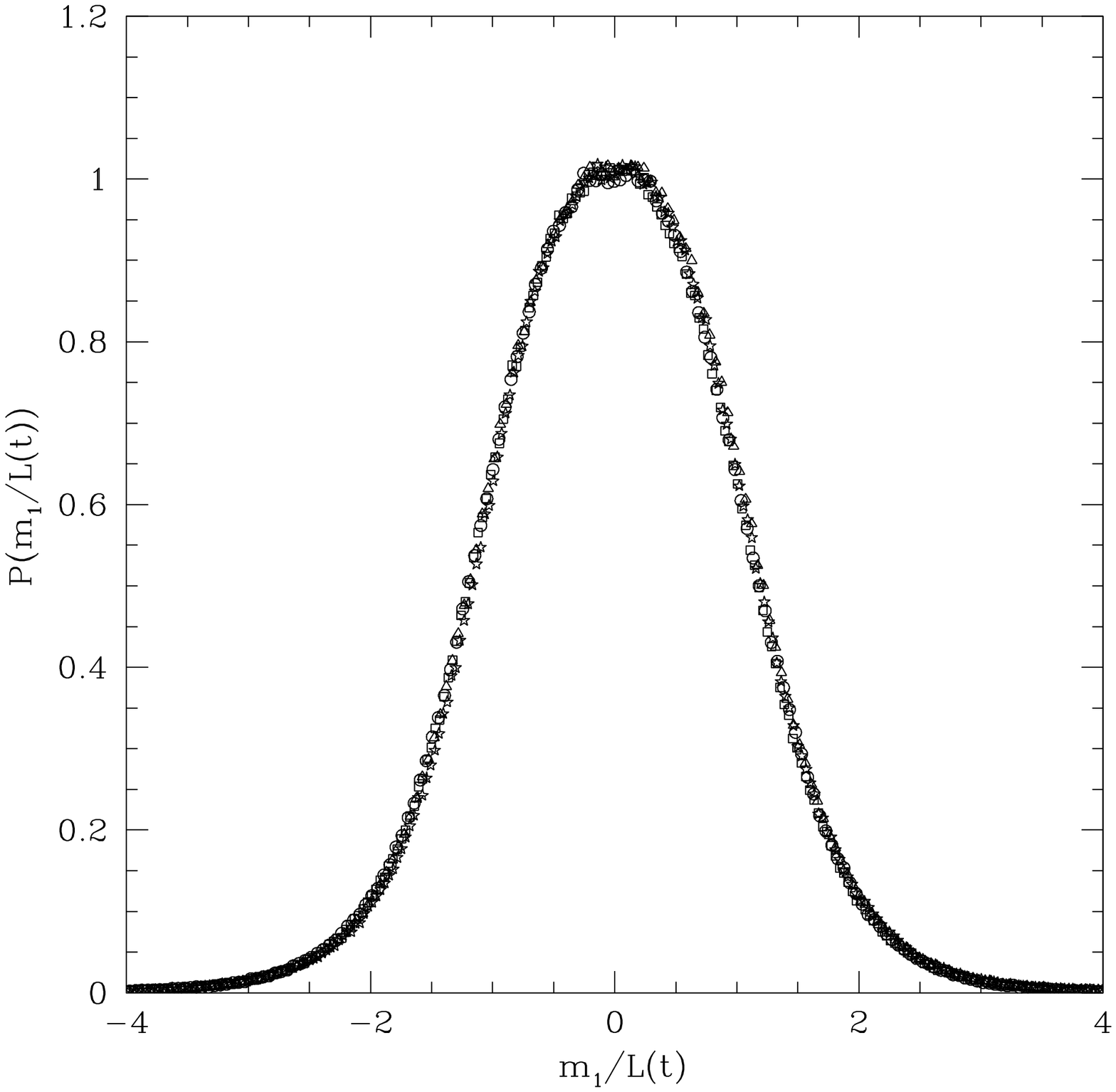,width=7.0cm,height=7.0cm}}
\end{figure}

Fig. 3.  Scaling plot of the un-normalised $P(m_1)$ for
$g=0(t=5000(\circ),\,t=10000(\Box))$ and
$g=1(t=5000(\triangle),\,t=10000(\ast))$. 

\vspace{1 cm}
 
\subsubsection{Conserved Case}

In the conserved case, the single point probability density
$P(m_1({\bf r},t))$ scales at late times, but the scaled distribution
is distinctly flatter than a gaussian (Fig.\ 4). Apart from this
strong deviation we find that the gaussian assumption leads to an
internal inconsistency, using a criterion initially developed by Yeung
et. al. \cite{CHUCK} for a conserved scalar order parameter. 

We numerically evaluate the spectral density, the fourier transform of
$\gamma(r,t)=\langle \vec m({\bf r}+{\bf x},t)\cdot\vec m({\bf x},t)
\rangle /\langle m^2({\bf r},t)\rangle$. We observe \cite{DAS2} that
the spectral density, which should be a strictly positive function of
its arguments, becomes negative for $k/k_{m} < 0.5 $ ($\gamma(k,t)$ is
peaked at $k_{m}$) and in the range $1.5<k/k_{m}<3.0$ ! This
demonstration highlights the intrinsic inconsistency of the gaussian
approach for conserved vector order parameters.  

\begin{figure}
\centerline{\epsfig{figure=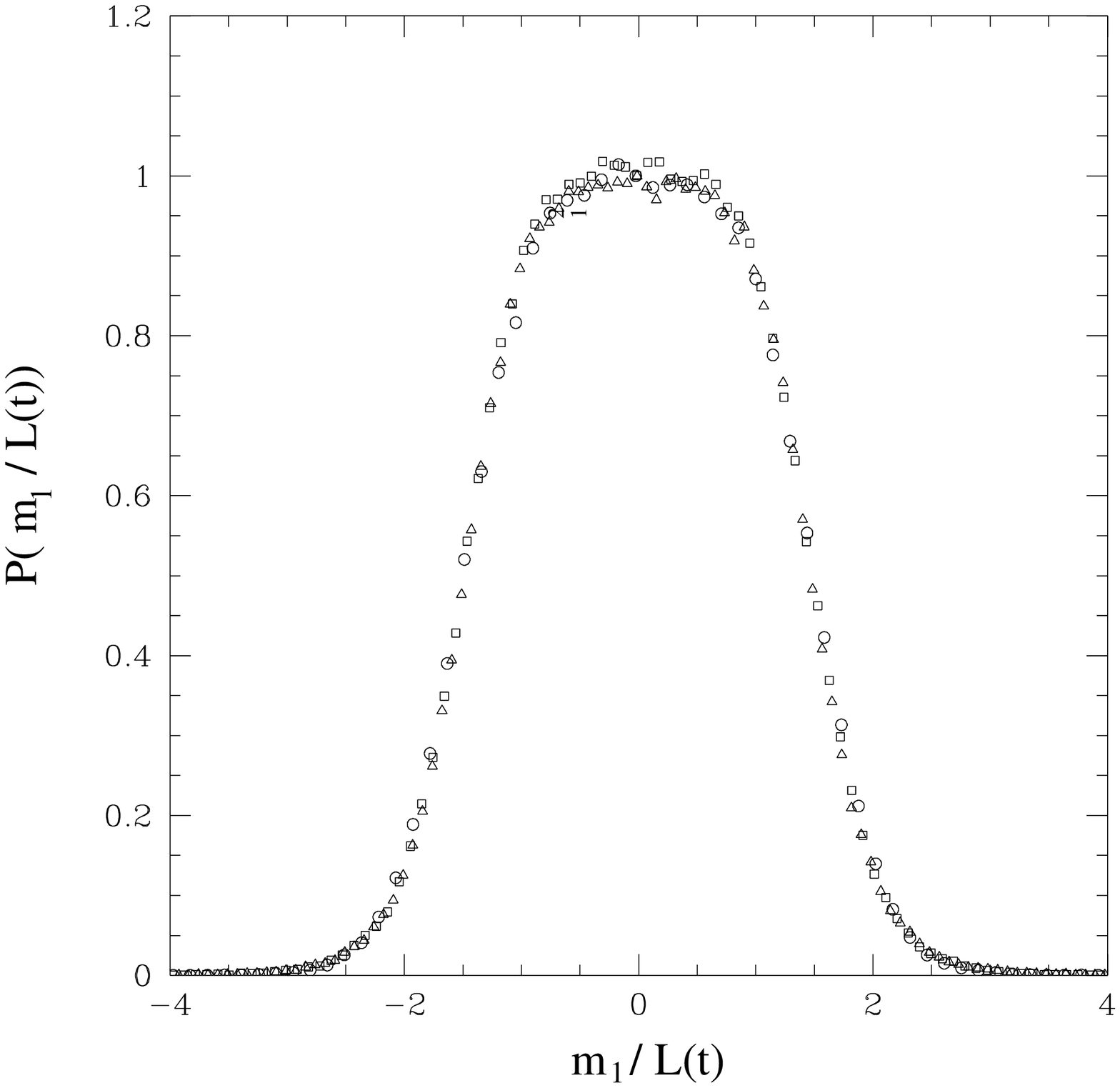,width=7.0cm,height=7.0cm}}
\end{figure}
 
Fig. 4. Scaling plot of the un-normalised $P(m_1)$ for $g=0$ at different
times $t = 900 (\circ), \, t = 3600(\Box),\, t = 6300(\triangle)$.

\vspace{.5 cm}

\subsection{ Multiscaling Analysis}

We have just seen that the Mazenko assumption of a gaussian
probability distribution fails sharply when the dynamics conserves the
total magnetisation. This seems to lead to an interesting technical
problem.  Coniglio and Zannetti \cite{CZ} had explicitly shown that
the late time dynamics of a conserved $n$-component spin model in the
limit $n \rightarrow \infty$ reveals an infinity of length scales
leading to a more complicated multiscaling form for the structure
factor (fourier transform of $C({\bf r},t)$) $S(k,t)\sim
L(t)^{p(k/k_m)d}$. The length scale $L(t)\sim t^{1/4}$ and the
position of the maximum of the structure factor $k_m^{-1}\sim (t/ \ln
t)^{1/4}$, and the scale dependent exponent $p(x)=1-(1-x^2)^2$. They
went on to speculate that this multiscaling behaviour might be a
generic feature of conserved order parameters \cite{CZ}. 

Subsequent numerical work showed that for the scalar ($n=1$)\cite{CO},
the XY ($n=2$)\cite{SR,MA} and $n=3,4$\cite{MA} models, the structure
factor obeyed the conventional `single length' scaling form. This
suggested the possibility  that multiscaling was nongeneric and an
`aberration' of the $n \rightarrow \infty$ model, though there was no
general proof. 

Bray and Humayun (BH) \cite{HUM} provided such a `proof'. Their analysis
was built on the validity of the Mazenko gaussian assumption for all
values of $n$.  Performing a $1/n$ expansion within the Mazenko framework,
they showed that the asymptotic $S({\bf k},t)$ exhibited multiscaling only
when $n=\infty$.

Having just demonstrated that the Mazenko gaussian assumption is
inconsistent for the conserved Heisenberg model ($n=3$), how do we
understand the BH proof \cite{HUM} ? Are the conclusions arrived by BH
incorrect ?
    
We now show that the structure factor $S({\bf k},t)$ of the conserved 
$n=3$ model does NOT obey multiscaling, consistent with the main
conclusion of BH. This is done using the method described in
\cite{SR}.

The method \cite{SR} demands a very accurate determination of $S({\bf
k},t)$.  Since the numerical evaluation of $S({\bf k},t)$ is subject to
large errors (especially at small $k$), we fit a function $C_{f}({\bf
r},t)$ to the computed $C({\bf r},t)$ and then calculate the fourier
transform $S_f({\bf k},t)$. The fitting function for $C_{f}({\bf r},t)$
has been taken as $\sin(r/L)/(r/L)(1+a(r/L)^2)\exp(-b(r/L)^2)$ which is
similar to the analytic form given in \cite{ROJAS}. Note that only $b$ and
$L$ are independent fitting parameters, $a$ is determined by the condition
$S_{f}(k=0,t)=0$. We now plot $S_f(k,t)$ versus $t$ at fixed values of
$x=k/k_m$ (Fig. 5). The resulting straight lines labeled by different
values of $x$, all show a constant slope of approximately $3/4$ (Fig. 5).
Using the proposed multiscaling form, a plot of $p(x)$ versus $x$ (inset
Fig. 5) shows that $p(x)$ is clustered around $1$. The small spread of
$p(x)$ around $1$ indicates that we have not quite reached the asymptotic
regime, and it is likely that the late time $p(x) \rightarrow 1$, in
agreement with conventional scaling. In addition, note that the form of
$p(x)$ is qualitatively different from the downward curving $p(x)$
predicted by Coniglio and Zannetti \cite{CZ}. We conclude then that the
correlation function $C({\bf r},t)$ for the $n=3$ conserved model does not
obey multiscaling.

\begin{figure}
\centerline{\epsfig{figure=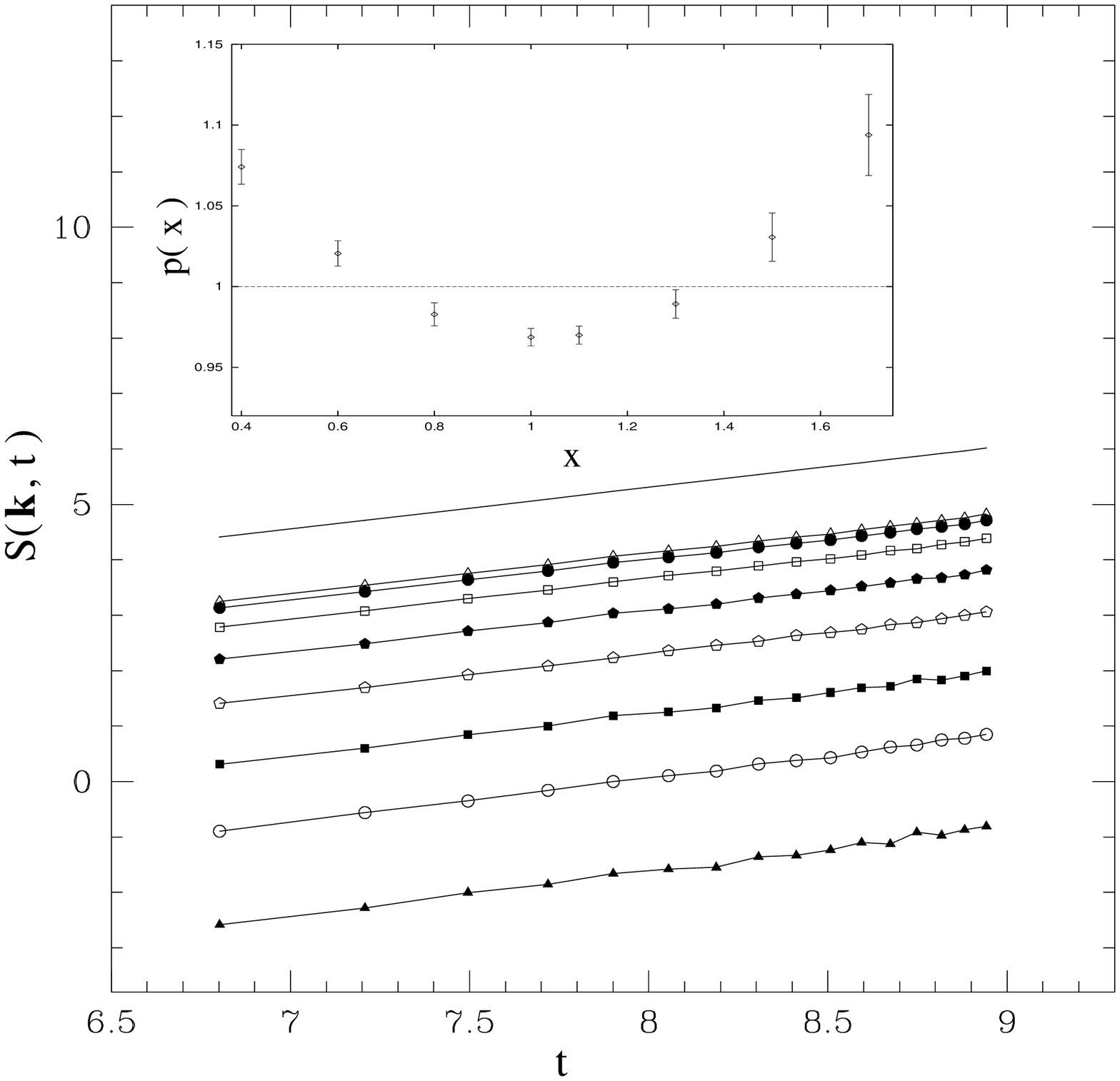,width=7.0cm,height=7.0cm}}
\end{figure}
 
Fig. 5. Plot of $S({\bf k},t)$ vs $t$ for $x=0.4$(circle),
$x=0.6$(pentagon), $x=0.8$(square), $x=1.0$(triangle), $x=1.1$(filled
circle), $x=1.3$(filled pentagon), $x=1.5$(filled square),
$x=1.7$(filled triangle). A straight line of slope $3/4$ is given for
comparison. Inset figure shows plot of $p(x)$ with $x$. Note that the
errors in $p(x)$ increases as $ \mid x -1 \mid $ increases because of
the smallness of $S({\bf k},t)$ near its wings.  

\vspace{.5 cm}

Thus we are forced to conclude that since the Mazenko assumption is
invalid for conserved vector order parameters (except when $n=\infty$), an
analytical proof of the absence of multiscaling of the asymptotic
structure factor is still lacking. 

\section{ Phase Ordering Dynamics\,: Quenches to $T = T_c$}

We have also studied the asymptotic dynamics following a quench to the
critical point \cite{DAS2}. We ask for whether the torque $g$ is relevant
at the Wilson-Fisher fixed point of the pure $g=0$ Heisenberg model, both
when the magnetisation is conserved and nonconserved.

Power counting reveals that $g$ is irrelevant for $d>2$ when the order
parameter is nonconserved. Thus the values of $z$ and $\lambda$ are
unchanged from the Wilson-Fisher value. For the conserved model, on the
other hand, power counting shows that $g$ is relevant for $d<6$, giving
rise to a new fixed point. The dynamical exponent $z$ changes from the WF
value of $4-\eta$ to $z=4-\varepsilon/2+{\cal O}(\epsilon^2)$, where
$\epsilon=4-d$ and $\varepsilon=6-d$ \cite{HAL}. The value of the 
autocorrelation exponent $\lambda$ is
however solely determined by the conservation law; it is always fixed at
the spatial dimension $d$. This is confirmed by a perturbative analysis to
all orders in $\epsilon$ \cite{DAS2}.


\bigskip

\begin{references}

\vspace{-1cm}

\bibitem[\dagger]{JAY} email\,: jayajit@rri.ernet.in

\bibitem[\star]{MAD} email\,: madan@rri.ernet.in\\
On leave of absence from\,:
Institute of Mathematical Sciences,
CIT Campus, Taramani, Chennai 600 113, India.


\bibitem{BRAY} A.\ J.\ Bray, Adv.\ Phys.\ {\bf 43} \ (1994) 357. 

\bibitem{DAS1} J.\ Das and M.\ Rao, Phys.\ Rev.\ E \ {\bf 57} \ (1998) 5069. 

\bibitem{DAS2} J.\ Das and M.\ Rao, IMSc. Preprint - 99/01/04. 

\bibitem{YRD}
C.\ Yeung, M.\ Rao, and R.\ C.\ Desai, Phys.\ Rev.\ E \ {\bf 53} \ (1996) 1.

\bibitem{BPT}
The computed $C(r,t)$ compares well with an approximate form (A.\
J.\ Bray and S.\ Puri, Phys.\ Rev.\ Lett.\ {\bf 67} \ (1991) 2670\,; H.\
Toyoki, Phys.\ Rev.\ B \ {\bf 45} \ (1992) 1965.) given by $f(x) =
(3\,\gamma/2\,\pi)\,[\,B(2,1/2)\,]^2\,_{2}F_{1}(1/2, 1/2, 5/2\,;
\gamma^2)$ where $\gamma = \exp(-x^2/8)$ and $B$ and ${}_{2}F_{1}$ are the
Beta and the hypergeometric functions respectively.



\bibitem{MAZENKO}
G.\ F.\ Mazenko, Phys.\ Rev.\ B \ {\bf 43} \ (1994) \ 357\,;
A.\ J. \ Bray and K.\ Humayun, J. Phys. A \ {\bf 25}, \ (1992) \ 2191.

 
\bibitem{CHUCK}
C.\ Yeung, A.\ Shinozaki and Y.\ Oono, Phys.\ Rev.\ E \ {\bf 49}\ (1994) \ 2693.

\bibitem{CZ}
A.\ Coniglio and M.\ Zannetti, Europhys.\ Lett.\ {bf 10} \ (1989) \ 57.


\bibitem{CO}
A.\ Conigloi,\ Y.\ Oono,\ A.\ Shinozaki,\ and M.\ Zannetti,\ Europhys. Lett.
\ {\bf 18} \ (1992) \ 59.


\bibitem{SR}
M.\ Seigert and M.\ Rao, Phys.\ Rev.\ Lett.\ {\bf 70} \ (1993) \ 1956.

\bibitem{MA}
M.\ Rao, and A.\ Chakrabarti, Phys.\ Rev.\ E {\bf 49} \ (1994) \ 3727.

\bibitem{HUM}
A.\ J.\ Bray and K.\ Humayun\ , Phys.\ Rev.\ Lett.\ {\bf 68} \ (1992) \ 1559.


\bibitem{ROJAS}
F.\ R.\ Iniguez and A.\ J.\ Bray, Phys.\ Rev.\ E \ {\bf 51} \ (1995) \ 188.

%
%
%
%
%
\bibitem{HAL}
P.\ C.\ Hohenberg and B.\ I.\ Halperin, Rev.\ Mod.\ Phys.\ {\bf 49}\ (1977)\ 436
\, ;  S.\ K.\ Ma and G.\ F.\ Mazenko, Phys.\ Rev.\ B {\bf 11} \ (1975)\ 4077.

\end{references}
\end{document}